\documentclass[a4paper]{jpconf}
\usepackage{graphicx}
\begin{document}
\title{From Cavity Electromechanics to Cavity Optomechanics}

\author{C. A. Regal and K. W. Lehnert}

\address{JILA, National Institute of Standards and Technology and the University of Colorado, and Department of Physics, University of Colorado, Boulder, Colorado 80309, USA}

\ead{regal@colorado.edu}

\begin{abstract}

We present an overview of experimental work to embed high-$Q$ mesoscopic mechanical oscillators in microwave and optical cavities.  Based upon recent progress, the prospect for a broad field of ``cavity quantum mechanics" is very real.  These systems introduce mesoscopic mechanical oscillators as a new quantum resource and also inherently couple their motion to photons throughout the electromagnetic spectrum.

\end{abstract}

Mechanical oscillators coupled to the electromagnetic mode of a cavity have emerged as an important new frontier in quantum optics.  By utilizing low-optical-loss and high-$Q$ nano- and micromechanical elements, researchers can now achieve significant coupling to the cavity mode compared to mechanical decoherence.  This coupling presents an opportunity for ground-state cooling of the mechanical oscillator and even strong many-photon coupling between the mechanical oscillator and cavity mode.  A unique feature of cavity mechanics is that the coupling does not rely on resonant behavior of the intracavity element.  Thus, in principle, the concept works for any frequency of light where a mechanical element can be integrated into a high-finesse cavity.  Indeed, experiments are proceeding in parallel throughout the electromagnetic spectrum.

\begin{figure}[h] \begin{center}
\includegraphics[width=350pt]{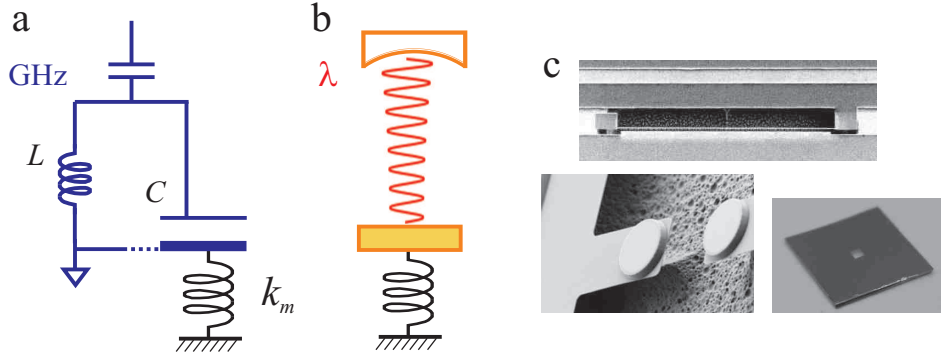}
\caption{\label{devices} (a) Schematic of an electromagnetic cavity (LC oscillator) linearly coupled to the motion of a mechanical oscillator with spring constant $k_m=m \omega_m^2$. (b) The optical analog.  A Fabry-Perot optical cavity with a moving end mirror.  (c) Examples of mechanical elements used in these experiments.  A metallic nanostring \cite{Regal2008}, a mirror on a cantilever \cite{Groblacher2009}, a stoichiometric silicon nitride membrane \cite{Wilson2009}.}
\end{center} \end{figure}

Near ground-state cooling has now been achieved with both metallic oscillators coupled to microwave superconducting resonators \cite{Teufel2008,Rocheleau2009} and mirrors coupled to high-finesse optical cavities \cite{Groblacher2009b,Schliesser2009,Park2009}.  While in one case the electromagnetic resonance is defined by an LC oscillator and in the other case by the distance between two mirrors (Fig. \ref{devices}), both employ the same parametric coupling.  The motion of the mechanical oscillator changes the cavity resonance according to a coupling coefficient $g$; the term in the Hamiltonian describing this radiation pressure interaction is $H_{\rm int} = \hbar g x_{\rm zp} a^\dag  a(d^\dag   + d)$.  Here $a$ represents the cavity mode, and the position operator of the mechanical membrane is $\hat x = x_{\rm zp} (d^\dag   + d)$, where $x_{\rm zp}$ is the mechanical oscillator's zero point motion.

In this article, we review some of the progress towards quantum control of cavity mechanical systems, addressing both electromechanical and optomechanical realizations.  Similar themes are found for the mechanical oscillators used for both electromechanics and optomechanics.  Thus, we can propose that a mechanical oscillator could be physically common to two cavities, one at an optical frequency and one at a microwave frequency.  Such an interface, in combination with strong coupling and slow decoherence, could allow for protocols such as quantum-state transfer between microwave and optical photons.

One of the first goals within the cavity mechanics community has been ground-state cooling of a mechanical oscillator by harnessing the radiation pressure within a cavity.  This cooling has been discussed in a number of review articles (see for example \cite{Kippenberg2007}) and would be an important demonstration of the quantum capabilities of cavity mechanics.  The idea is analogous to resolved sideband cooling used for many years in atomic physics experiments \cite{Wineland1979}, with the difference that the dissipation is provided by photons leaking from the cavity rather than by spontaneous emission.  This idea of cavity cooling was also demonstrated in experiments with single atoms and atomic gases in cavities \cite{Maunz2004,Leibrandt2009,Brennecke2008,Murch2008}.

\begin{figure}[h] \begin{center}
\includegraphics[width=270pt]{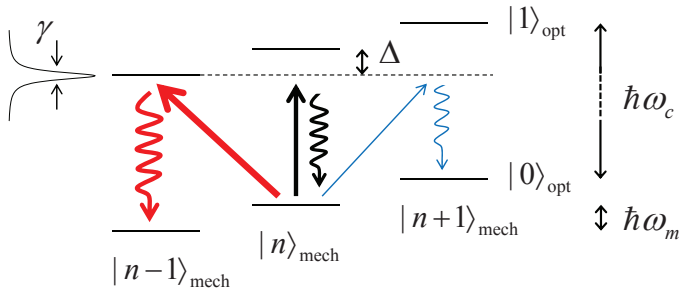}
\begin{minipage}[b]{14pc}\caption{\label{cooling} Ladder of mechanical oscillator motional states demonstrating resolved-sideband cavity cooling.  The excitation laser is red-detuned of the cavity transition, $\omega_c$, and the dominant transition is indicated by the red arrows on the left.  This absorption followed by emission of a higher-energy photon gives rise to mechanical cooling.}
\end{minipage}
\end{center}
\end{figure}

As illustrated in Fig. \ref{cooling}, a sufficiently long cavity lifetime allows one to resolve individual motional states, i.e., achieve the resolved sideband limit where the mechanical frequency is larger than the cavity linewidth $\omega_m \gg \gamma$ \cite{Marquardt2007a,WilsonRae2007}.  To achieve ground-state cooling, one must overcome the heating from the environment.  The mechanical oscillator is physically coupled to a solid-state object at a temperature $T_{\rm bath}$ and only isolated via its mechanical quality factor $Q_m=\omega_m/\gamma_m$.  Its heating rate $\gamma_m n_{\rm bath}$, where $n_{\rm bath}=k_B T_{\rm bath}/\hbar \omega_m$, must be overcome by the cooling rate.   At the optimum red detuning of $\omega_c-\omega_0=\Delta=\omega_m$ and in the fully resolved sideband limit, the cooling rate is $\Gamma_{\rm cool}=4 \Gamma^2/\gamma$, where $\Gamma=g x_{\rm zp} |\alpha|$ is the effective coupling with $|\alpha|^2$ photons in the cavity.  The ratio of these parameters $\gamma_m n_{\rm bath}/\Gamma_{\rm cool}$ is the final phonon occupation.  This analysis is valid in the limit where $\Gamma$ and $\gamma_m n_{\rm bath}$ remain small compared to other parameters in the problem.  For example, an important minimum condition is $\gamma_m n_{\rm bath} \ll \omega_m$, which is equivalent to a requirement on the $Q$-frequency product of $Q_m \omega_m \gg k_B T_{\rm bath}/\hbar$.  This puts a stringent condition on the mechanical oscillators and, for the most part, has led to a focus on cryogenic precooling of cavity mechanics experiments.

Microwave and optical systems have different advantages and disadvantages for realizing cavity cooling to the ground state.  Optical systems appear at first glance to be the most natural for harnessing radiation pressure in cavities.  The coupling can naturally be quite large; for a moving end mirror the coupling is given by $g=\omega_c/L$.  Hence, for 1 $\mu$m light and a mm-length cavity, $g/2\pi \sim 300 \ {\rm kHz/pm}$.  Although the mirrors typically used in optical cavities are too massive to be strongly influenced by radiation pressure, advances in the microfabrication of optical quality elements have allowed rapid progress.  For example, experimenters have used microcavity technology \cite{Schliesser2009,Park2009} and placed a nm-thick mechanical membrane inside of a traditional high-finesse Fabry-Perot cavity \cite{Thompson2007}.  The challenge of optomechanics, however, is its compatibility with cryogenic operation.  As mentioned above, with typical $Q$-frequency products of mechanical oscillators, a ``quantum-enabled" system with $Q_m\omega_m > k_B T_{\rm bath}/\hbar$ is most likely to be realized with cryogenic precooling.  However, alignment of high-finesse cavities at cryogenic temperatures is tricky, and absorption of optical power that heats $T_{\rm bath}$ must be avoided by use of low-loss optical materials.  Despite these challenges, a number of experiments using a combination of cryogenic and cavity cooling have now reached the level of tens of phonons occupation in a mechanical oscillator.

Superconducting microwave cavities, on the other hand, are naturally compatible with a dilution refrigerator environment.  This compatibility provides a large advantage for the amount of cavity cooling required.  For example, starting at 50 mK with a 10 MHz mechanical oscillator leaves only $n_{\rm bath}=100$.  (In fact by using a GHz mechanical oscillator, ground-state cooling has recently been achieved with only conventional refrigeration techniques \cite{Oconnell2010}.)  With a traditional microwave cavity at the cm scale, however, the coupling to a micromechanical object would be very small.  Utilizing the one-dimensional (1D) geometry of a superconducting stripline cavity helped realize sufficient coupling to demonstrate microwave cavity cooling of cold oscillators \cite{Teufel2008a,Teufel2008,Rocheleau2009}, but increasing the coupling still remained a challenge.  The most straightforward, but difficult, way to improve the coupling is to squeeze the capacitor plates incorporating the mechanical element closer and closer together.  Another route is engineering the impedance $Z=\sqrt{L/C}$.  This route allows the capacitance from the mechanical object to be the dominant capacitance while maintaining GHz resonant frequencies \cite{Teufel2009}.  Recent experiments advancing both of these goals have allowed for a very large coupling $g/2\pi \sim 50 \ {\rm kHz/pm}$ \cite{Teufel2010} that is surprisingly comparable to optomechanical systems and should allow cooling to very low occupation \cite{Teufel2010b}.

\begin{figure}[h] \begin{center}
\includegraphics[width=350pt]{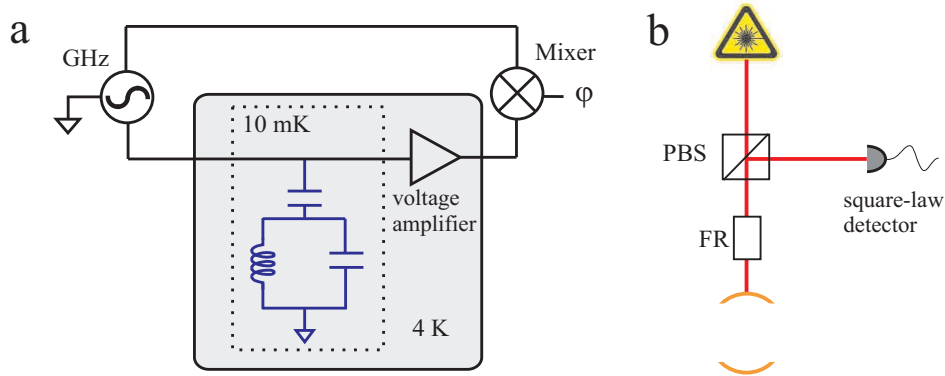}
\caption{\label{interferometer} Sketches of interferometers for motion detection.  In experiments thus far, the cavity is used as a transducer to measure the mechanical fluctuations.  (a) Microwave interferometer.  A typical cryogenic voltage amplifier is a high electron mobility transistor (HEMT) amplifier.  Recent experiments have incorporated a chain of both a Josephson parametric amplifier (JPA) and a HEMT amplifier (see text).   (b) Optical interferometer incorporating a Fabry-Perot cavity and a Pound-Drever-Hall scheme.  Only the most basic elements are depicted.}
\end{center} \end{figure}

Another important difference lies in detection.  An optical interferometer offers a well-tested way of measuring the mechanical motion near the quantum limit, while in electromechanical systems, the technology is just being developed.  Figure \ref{interferometer}(a) illustrates a typical setup for interferometric measurement of the mechanical motion in the microwave domain.  A key component is the cold microwave amplifier; the best available commercial amplifiers used in early experiments added 30 quanta of noise \cite{Regal2008}.  With these added quanta, the detection is equivalent to a square-law detector with a quantum efficiency of only $\eta=(1+2n_{\rm add})^{-1}\sim 2\%$.  To allow measurement of motion near the quantum ground state in a feasible amount of time, new amplifier technology needed to be developed. In fact tunable Josephson parametric amplifiers (JPA) capable of detecting microwave fields with an added noise of less than half a quantum have recently been implemented \cite{Castellanos2008, Teufel2009}.

The mechanical elements used in both electromechanical and optomechanical experiments draw on a large variety of high-$Q$ oscillators.  Their masses range from the gram-scale mirrors that have been utilized by the LIGO collaboration \cite{Corbitt2007} to picogram masses of 100 nm cross-section beams \cite{Regal2008,Anetsberger2010}.  For the micro- and nanomechanical elements, our field draws heavily on the success of micro- and nanoelectromechanical systems (MEMS and NEMS) \cite{Ekinci2005a}.  Extensive work to understand and develop high-quality factor, low-mass oscillators in the 1990s resulted in a variety of sensitive mass and force detectors.  In the quest to realize quantum cavity mechanics, we similarly want low mass and high quality factor, but we also want to optimize upon large absolute mechanical frequency compared to $T_{\rm bath}$.  This optimization on large frequency is not required for a force detector where the ultimate sensitivity is limited by the thermal force noise of spectral density $S_F=4m\gamma_m k_BT_{\rm bath}$, which depends on the properties of the mechanical oscillator only through $m$ and $\gamma_m$.  Thus, in both optomechanics and electromechanics we have gravitated towards certain classes of mechanical objects that suit our particular needs.

\begin{figure}[h] \begin{center}
\includegraphics[width=305pt]{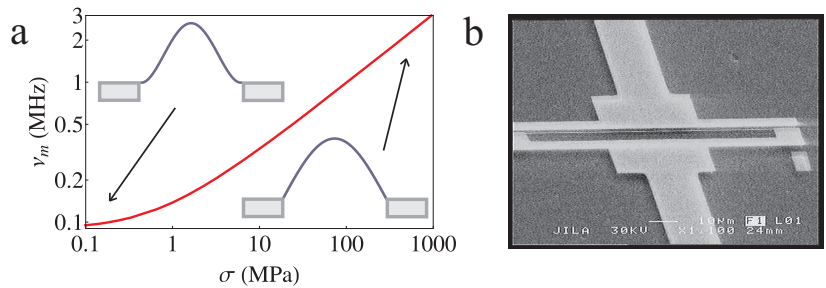}
\caption{\label{strings} (a) Variation of resonant frequency with film stress $\sigma$ for a beam with clamped-clamped boundary conditions \cite{Bokaian1990,Unterreithmeier2010}.  The typical string considered here has cross-sectional dimensions of 100 nm and is 100 $\mu$m long. Insets:  Illustration of how the fundamental mode shape (exaggerated) changes from that of a flexural mode to near a string mode.    (b) Example of a metallic nanostring placed between two electrodes [picture credit P.-L. Yu].}
\end{center}
\end{figure}

One example is tensioned mechanical oscillators (Fig. \ref{strings}).  Much of the MEMS and NEMS field is based upon flexural modes of nanobeams, but if one adds tension to a 1D or 2D oscillator to make a string or drum the mechanical frequency can be significantly increased.  In addition the quality factor increases, allowing for a double benefit in the $Q$-frequency product.  Tensioned oscillators have been used in electromechanical devices in the form of polycrystalline metallic nanostrings \cite{Regal2008}, metallized silicon nitride (SiN) strings \cite{Hertzberg2010}, and metallic drums \cite{Teufel2010}.  SiN has been used for optomechanics in the form of both drums and strings \cite{Thompson2007, Wilson2009, Anetsberger2010}.  The optical spring effect used in the LIGO collaboration also uses a similar idea of increasing stiffness without increasing loss \cite{Corbitt2007}.  When the stress of SiN film is pushed to its limit (approaching 1 GPa) by using the stoichiometric form, ${\rm Si_3N_4}$, the $Q$-frequency products, even at room temperature, can be quite spectacular \cite{Verbridge2006,Verbridge2007,Southworth2009}.  In fact, they can exceed $k_B T_{\rm room}/h=6\times10^{12}$  Hz and be incorporated into low-loss cavity mechanics systems \cite{Wilson2009}.  This extends the prospects of quantum cavity mechanics to the room temperature domain.

Improving upon this encouraging result requires that we understand the dependence of the quality factor on tension in these resonators.  The gross behavior of a significant increase in $Q_m$ is apparent; mainly this effect can be understood simply as increasing the resonance frequency while keeping the dissipation, as expressed through the linewidth $\gamma_m$, approximately constant.  For the case of SiN strings with loss dominated by coupling to uniformly-distributed localized defects, this was verified quantitatively in the experiments of Ref. \cite{Unterreithmeier2010}. The bending energy, associated with the loss, was calculated and observed to change much less than the total energy.  However, the model showed some change in the linewidth with tension.  This work used SiN strings of varying lengths and mode index, but it will be interesting to investigate the loss mechanisms of tensioned resonators of different materials and different geometries.  A variety of loss mechanisms considered extensively for flexural modes require reanalysis for tensioned resonators.  For example, $Q$ limits due to thermoelastic loss are expected to be surprisingly large \cite{Zwickl2007}, and anchor or clamping losses will need to be considered quantitatively in the presence of tension.

Thus far we have considered the goal of initializing a mechanical oscillator in the ground state via cavity cooling.  While this is an important indication that the system is quantum enabled, one larger goal is to determine the most appropriate role of cavity quantum mechanics within a quantum network \cite{Kimble2008}.  We suspect cavity mechanics can play a unique role in quantum information via the ability to facilitate wavelength conversion.  In particular, one can envision the mechanical oscillator as a mediator for quantum-state transfer between photons in two different cavities to which the mechanical oscillator is simultaneously coupled  \cite{Tian2010}.  Such controlled quantum protocols require reversible energy exchange to take place faster than loss of energy to the environment.  This regime of strong coupling is routinely achieved for various atomic and solid state qubits and photons.  For cavity mechanical systems, the single-photon coupling rate is typically in the range of $g x_{\rm zp}/2\pi\sim10$ Hz, which is much smaller than other rates in micromechanical cavity systems, which are typically kHz$-$MHz.  However, it is possible to achieve a regime of strong many-photon coupling \cite{Marquardt2007a,Dobrindt}, as pioneered in the work in Ref. \cite{Groblacher2009}.  Here they realized a coupling $\Gamma=g x_{\rm zp} |\alpha|$ greater than both $\gamma$ and $\gamma_m$. While the ability to achieve only many-photon strong coupling precludes protocols that require nonlinearities, there are many things one can imagine doing, including quantum-state transfer, with the linear coupling.

It is possible cavity mechanics could even be useful to convert microwave and optical photons, a wavelength conversion problem that has attracted more and more interest \cite{Andre2006,Stannigel2010}.  Superconducting Josephson junction-based qubits and superconducting resonant cavities have emerged as the ideal realization of quantum two-level systems interacting with a single mode of the electromagnetic spectrum. Remarkably, it is now possible to deterministically create arbitrary states of that single mode containing up to 15 photons by controllable interaction with a two-level system \cite{Hofheinz2009}. The single mode is a lithographically fabricated electrical circuit resonant at microwave frequencies. The superconducting circuitry provides a highly flexible platform for creating on-demand complex quantum states of a light field.  Unfortunately, microwaves can only be distributed efficiently over superconducting cable.  Thus, the microwave quantum states are essentially stuck inside of the dilution refrigerator where they are created, and further they cannot be preserved for times longer than $\sim 10$ $\mu$s.  If, however, it were possible to convert these states to optical wavelengths, they could be transmitted over kilometer distances with negligible loss.  They could then be stored for long times in, for instance, atomic ensembles \cite{Eisaman2005}.

\begin{figure}[h] \begin{center}
\includegraphics[width=140pt]{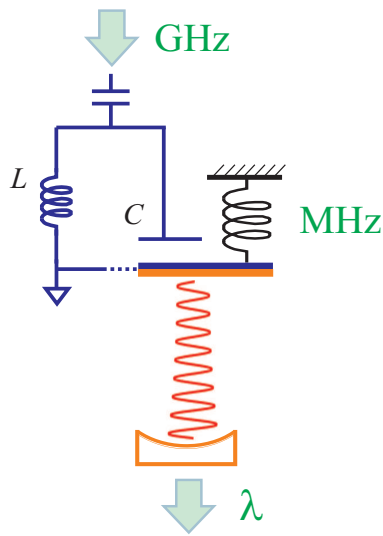}
\begin{minipage}[b]{14pc}\caption{\label{hybrid} The ability to realize both cavity electromechanics and optomechanics allows
one to envision hybrid electro-optomechanical devices.  Here the position of a mechanical oscillator changes simultaneously the
resonant frequencies of both the microwave and optical resonances.  Note no frequency matching is required; the mechanical oscillator in the MHz regime acts as a mediator between the microwave and optical photons.}
\end{minipage}
\end{center}
\end{figure}

Explicitly, the position of the mechanical oscillator could tune the resonance frequency of two cavities, one microwave and one optical (Fig. \ref{hybrid}).  The device is described by the following Hamiltonian
\begin{equation}\label{Eq:SystemHamiltonian}
H = H_0  + H_E  + H_M  + \hbar g_{OM} x_{\rm zp} a^\dag  a(d^\dag   + d) - \hbar g_{EM} x_{\rm zp} b^\dag  b(d^\dag   + d)
\end{equation}
where $H_0$, $H_{E}$, and $H_M$ are the optical, electrical, and mechanical harmonic oscillator Hamiltonians; the two subsequent terms represent the optomechanical ($OM$) and electromechanical ($EM$) interaction.  The crucial point for the goal of state transfer is that by detuning a drive tone red of the cavity resonance, just as in cavity cooling, a beamsplitter interaction can be induced \cite{Zhang2003,Akram2010}.  With a cavity driven red detuned at $\Delta = \omega_m$ in the resolved-sideband limit, the coupling in the interaction representation is $\hbar \Gamma_{OM} (\tilde d^\dag  \tilde a + \tilde a^\dag  \tilde d)$, where $\Gamma_{OM}=g_{OM} x_{\rm zp} |\alpha|$.  An equivalent beamsplitter interaction is found for the microwave cavity. Thus, one imagines that by appropriately strongly driving both the optical and microwave cavities red-detuned, we can swap microwave and optical photons that have very different absolute frequencies, but the same frequency difference relative to their respective drives.

\begin{figure}[h] \begin{center}
\includegraphics[width=370pt]{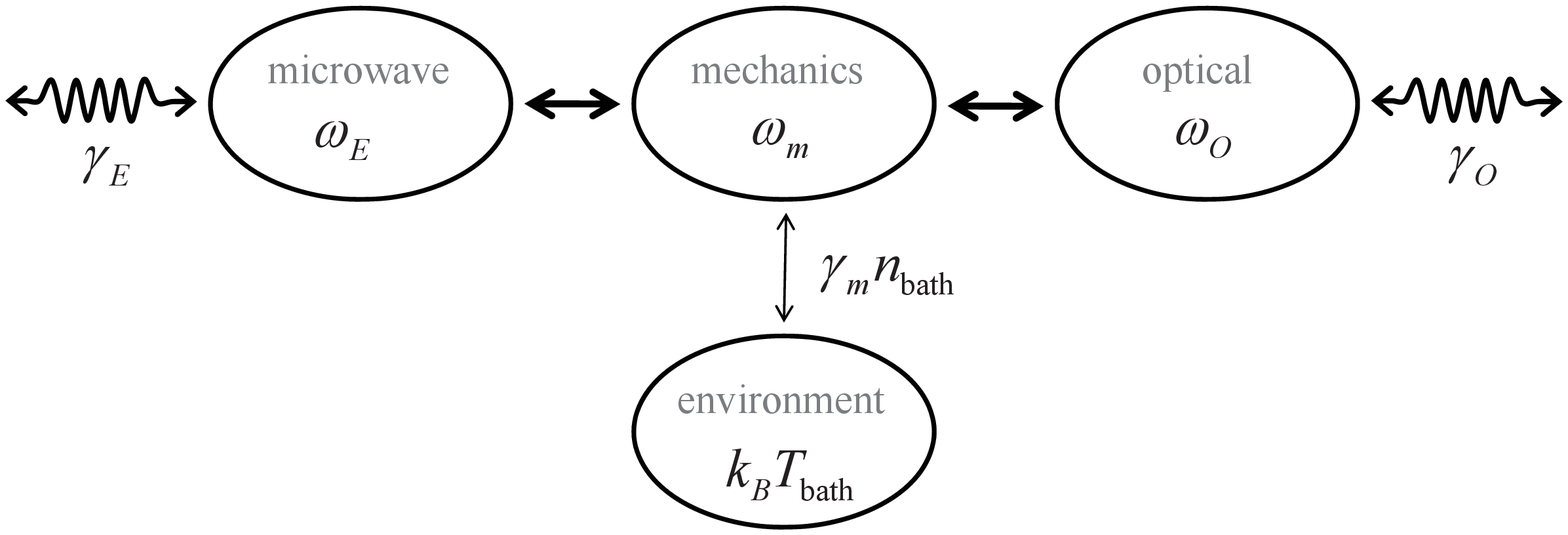}
\caption{\label{decay} Channels for coupling and loss that are relevant to mechanically mediated quantum-state transfer
between photons in two cavities.  Here we consider the case of coupling between microwave, $E$, and optical, $O$, cavities.}
\end{center}
\end{figure}

For coherent exchange of optical and microwave photons, the critical question is whether these swap interactions can be accomplished before decoherence processes destroy the quantum state.  Figure \ref{decay} demonstrates the various rates that must be considered to answer this question.  Very generally, the goal is to achieve strong coupling rates $\Gamma$ compared to cavity decay $\gamma$ or mechanical decoherence $\gamma_m n_{\rm bath}$, without exceeding $\omega_m/2$ where bistability occurs.  On the practical side, there are a variety of questions to answer about the feasibility of creating the device shown in Fig. \ref{hybrid}.  While mechanical oscillators can be very versatile elements, it is not a priori clear that a common mechanical oscillator design can work as both a capacitor plate and a mirror. However, as discussed above, there do appear to be materials that could allow for such a hybrid element.  In particular, tensioned-drum resonators appear to be good candidates for both electromechanics and optomechanics, and one could envision separating the metallic and dielectric parts of the drum spatially using higher-order mechanical modes.

The quest to control the motion of mesoscopic oscillators is still in its infancy compared to abilities of atomic systems such as trapped ions.  Yet, the versatile nature of mechanical oscillators and the inherent coupling to the electromagnetic field that cavity mechanics provides may allow this new system to play an important role in future quantum protocols.

\noindent {\bf Acknowledgements} \newline
We thank J. D. Teufel, R. W. Simmonds, T. Donner, H. J. Kimble, S. B. Papp, D. J. Wilson, A. A. Clerk, and J. Gambetta for helpful discussions.  We acknowledge support from the Clare Boothe Luce Foundation, DARPA-QUASAR, NSF, and NIST.

\section*{References}

\bibliographystyle{iopart-num}
\bibliography{icap}

\providecommand{\newblock}{}
\begin{thebibliography}{10}
\expandafter\ifx\csname url\endcsname\relax
  \def\url#1{{\tt #1}}\fi
\expandafter\ifx\csname urlprefix\endcsname\relax\def\urlprefix{URL }\fi
\providecommand{\eprint}[2][]{\url{#2}}

\bibitem{Teufel2008}
Teufel J~D, Harlow J~W, Regal C~A and Lehnert K~W 2008 {\em Phys. Rev. Lett.\/}
  {\bf 101} 197203

\bibitem{Rocheleau2009}
Rocheleau T, Ndukum T, Macklin C, Hertzberg J~B, Clerk A~A and Schwab K~C 2009
  {\em Nature\/} {\bf 463} 72

\bibitem{Groblacher2009b}
Gr\"{o}blacher S, Hertzberg J~B, Vanner M~R, Cole G~D, Gigan S, Schwab K~C and
  Aspelmeyer M 2009 {\em Nature Phys.\/} {\bf 5} 285

\bibitem{Schliesser2009}
Schliesser A, Arcizet O, Rivi\`ere R, Anetsberger G and Kippenberg T~J 2009
  {\em Nature Phys.\/} {\bf 5} 509

\bibitem{Park2009}
Park Y~S and Wang H 2009 {\em Nature Phys.\/} {\bf 5} 489

\bibitem{Regal2008}
Regal C~A, Teufel J~D and Lehnert K~W 2008 {\em Nature Phys.\/} {\bf 4} 555

\bibitem{Groblacher2009}
Gr\"{o}blacher S, Hammerer K, Vanner M~R and Aspelmeyer M 2009 {\em Nature\/}
  {\bf 460} 724

\bibitem{Wilson2009}
Wilson D~J, Regal C~A, Papp S~B and Kimble H~J 2009 {\em Phys. Rev. Lett.\/}
  {\bf 103} 207204

\bibitem{Kippenberg2007}
Kippenberg T and Vahala K 2007 {\em Opt. Exp.\/} {\bf 15} 17172

\bibitem{Wineland1979}
Wineland D~J and Itano W~M 1979 {\em Phys. Rev. A\/} {\bf 20} 1521--1540

\bibitem{Maunz2004}
Maunz P, Puppe T, Schuster I, Syassen N, Pinkse P~W~H and Rempe G 2004 {\em
  Nature\/} {\bf 50} 428

\bibitem{Leibrandt2009}
Leibrandt D~R, Labaziewicz J, Vuletic V and Chuang I~L 2009 {\em Phys. Rev.
  Lett.\/} {\bf 103} 103001

\bibitem{Brennecke2008}
Brennecke F, Ritter S, Donner T and Esslinger T 2008 {\em Science\/} {\bf 322}
  235

\bibitem{Murch2008}
Murch K~W, Moore K~L, Gupta S and Stamper-Kurn D~M 2008 {\em Nature Phys.\/}
  {\bf 4} 561

\bibitem{Marquardt2007a}
Marquardt F, Chen J~P, Clerk A~A and Girvin S~M 2007 {\em Phys. Rev. Lett.\/}
  {\bf 99} 093902

\bibitem{WilsonRae2007}
Wilson-Rae I, Nooshi N, Zwerger W and Kippenberg T~J 2007 {\em Phys. Rev.
  Lett.\/} {\bf 99} 093901

\bibitem{Thompson2007}
Thompson J~D, Zwickl B~M, Jayich A~M, Marquardt F, Girvin S and Harris J~G~E
  2008 {\em Nature\/} {\bf 452} 72--75

\bibitem{Oconnell2010}
O$'$Connell A~D, Hofheinz M, Ansmann M, Bialczak R~C, Lenander M, Lucero E,
  Neeley M, Sank D, Wang H, Weides M, Wenner J, Martinis J~M and Cleland A~N
  2010 {\em Nature\/} {\bf 464} 697

\bibitem{Teufel2008a}
Teufel J~D, Regal C~A and Lehnert K~W 2008 {\em New J. Phys.\/} {\bf 10} 095002

\bibitem{Teufel2009}
Teufel J~D, Donner T, Castesllanos-Beltra M~A, Harlow J~W and Lehnert K~W 2009
  {\em Nature Nanotech\/} {\bf 4} 820

\bibitem{Teufel2010}
Teufel J~D and Simmonds R~W 2010 {\em submitted\/}

\bibitem{Teufel2010b}
Teufel J~D, Donner T, Simmonds R~W and Lehnert K~W 2010 {\em in preparation\/}

\bibitem{Castellanos2008}
Castellanos-Beltran M~A, Irwin K~D, Hilton G~C, Vale L~R and Lehnert K~W 2008
  {\em Nature Phys.\/} {\bf 4} 929

\bibitem{Corbitt2007}
Corbitt T, Chen Y, Mueller-Ebhardt H, Innerhofer E, Ottaway D, Rehbein H, Sigg
  D, Whitcomb S, Wipf C and Mavalvala N 2007 {\em Phys. Rev. Lett.\/} {\bf 98}
  150802

\bibitem{Anetsberger2010}
Anetsberger G, Gavartin E, Arcizet O, Unterreithmeier Q~P, Weig E~M, Gorodetsky
  M~L, Kotthaus J~P and Kippenberg T~J 2010 {\em arXiv:1003.3752v1\/}

\bibitem{Ekinci2005a}
Ekinci K~L and Roukes M~L 2005 {\em Rev. Sci. Instrum.\/} {\bf 76} 061101

\bibitem{Hertzberg2010}
Hertzberg J~B, Rocheleau T, Ndukum T, Savva M, Clerk A~A and Schwab K~C 2010
  {\em Nature Phys.\/} {\bf 6} 213

\bibitem{Verbridge2006}
Verbridge S~S, Parpia J~M, Reichenbach R~B, Bellan L~M and Craighead H~G 2006
  {\em J. Appl. Phys.\/} {\bf 99} 124304

\bibitem{Verbridge2007}
Verbridge S~S, Shapiro D~F, Craighead H~G and Parpia J~M 2007 {\em Nano
  Lett.\/} {\bf 7} 1728

\bibitem{Southworth2009}
Southworth D~R, Barton R~A, Verbridge S~S, Ilic B, Fefferman A~D, Craighead H~G
  and Parpia J~M 2009 {\em Phys. Rev. Lett.\/} {\bf 102} 225503

\bibitem{Unterreithmeier2010}
Unterreithmeier Q~P, Faust T and Kotthaus J~P 2010 {\em Phys. Rev. Lett.\/}
  {\bf 105} 027205

\bibitem{Zwickl2007}
Zwickl B~M, Shanks W~E, Jayich A~M, Yang C, Bleszynski A~C, Thompson J~D and
  Harris J~G~E 2007 {\em Appl. Phys. Lett.\/} {\bf 92} 103125

\bibitem{Bokaian1990}
Bokaian A 1990 {\em J. Sound and Vib.\/} {\bf 142} 481

\bibitem{Kimble2008}
Kimble H~J 2008 {\em Nature\/} {\bf 453} 1023

\bibitem{Tian2010}
Tian L and Wang H 2010 {\em arXiv:1007.1687v1\/}

\bibitem{Dobrindt}
Dobrindt J~M, Wilson-Rae I and Kippenberg T~J 2008 {\em Phys. Rev. Lett.\/}
  {\bf 101} 263602

\bibitem{Andre2006}
Andr\'e A, DeMille D, Doyle J~M, Lukin M~D, Maxwell S~E, Rabl P, Schoelkopf R~J
  and Zoller P 2006 {\em Nature Phys.\/} {\bf 2} 636

\bibitem{Stannigel2010}
Stannigel K, Rabl P, Sorensen A~S, Zoller P and Lukin M~D 2010 {\em
  arXiv:1006.4361v1\/}

\bibitem{Hofheinz2009}
Hofheinz M, and Ansmann H~W, Bialczak R~C, Lucero E, Neeley M, O'Connell A~D,
  Sank D, Wenner J, Martinis J~M and Cleland A~N 2009 {\em Nature\/} {\bf 459}
  546

\bibitem{Eisaman2005}
Eisaman M~D, Andr\'e A, Massou F, Fleischhauer M, Zibrov A~S and Lukin M~D 2005
  {\em Nature\/} {\bf 438} 837

\bibitem{Zhang2003}
Zhang J, Peng K and Braunstein S~L 2003 {\em Phys. Rev. A\/} {\bf 68} 013808

\bibitem{Akram2010}
Akram U, Kiesel N, Aspelmeyer M and Milburn G~J 2010 {\em New J. Phys.\/} {\bf
  12} 083030

\end{thebibliography}

\end{document}